\documentclass[dvips]{arxstspdf}
\usepackage{flushend}
\volume{25}
\issue{2}
\pubyear{2010}
\firstpage{158}
\lastpage{161}
\doi{10.1214/10-STS308A}
\referstodoi{10.1214/09-STS308}

\begin{document}
\begin{frontmatter}

\vspace*{6pt}
\title{Comment: The Need for Syncretism in Applied Statistics}
\runtitle{Comment}

\begin{aug}
\author{\fnms{Sander} \snm{Greenland}\ead[label=e1,text=lesdomes@ucla.edu]{lesdomes@ucla.edu}\corref{}}
\runauthor{S. Greenland}

\affiliation{University of California}

\address{S. Greenland is Professor,
Department of Epidemiology and Department of Statistics,
University of California,
Los Angeles, California 90095-1772, USA
 \printead{e1}}

\end{aug}

% ABSTRACT

% KEYWORDS

\end{frontmatter}

It is an honor to comment on Prof. Efron's latest contribution to the
merging of frequentist and Bayesian thinking into a harmonious (even if
not strictly coherent) statistical viewpoint. I will review my thinking
along those lines and some inspirations for it. I~agree with most of Dr.
Efron's views expressed here and in Efron (\citeyear{Efron2005}), with these important
exceptions: First, I disagree that frequentism has supplied a good set
of working rules. Instead, I argue that frequentism has been a prime
source of reckless overconfidence in many fields (especially but not
only in the form of 0.05-level testing; see Rothman, Greenland and Lash,
\citeyear{RothmanGreenlandLash2008}, Chapter~10 for
examples and further citations). I~also disagree that Bayesians are more
aggressive than frequentists in modeling. The most aggressive modeling
is that which fixes unknown parameters at some known constant like zero
(whence they disappear from the model and are forgotten), thus
generating overconfident inferences and an illusion of simplicity; such
practice is a hallmark of conventional frequentist applications in
observational studies.

As working rules, the problem with conventional methods lies not so much
with frequentism, but\break rather with frequentist tools for designed
experiments being misapplied to observational data\break (Greenland, \citeyear{Greenland2005a}).
Bayesians can and do misapply their methods similarly; they just haven't
been given as much opportunity to do so. Conversely, many frequentist as
well as Bayesian tools for observational studies have been developed,
especially for sensitivity analysis. But the overconfidence problem has
been perpetuated by the ongoing concealment of unbelievable point-mass
priors within models in order to maintain frequentist identification of
target parameters.

The problem can be addressed by sacrificing identification and replacing
bad modeling assumptions with explicit and reasonable priors (Gustafson, \citeyear{Gustafson2005};
Greenland, \citeyear{Greenland2005a,Greenland2009a}). Perhaps ironically, frequentist thought
experiments and simulations can then provide both contextual and
frequentist diagnostics (Rubin, \citeyear{Rubin1984}; Greenland, \citeyear{Greenland2006};
Gustafson and Greenland, \citeyear{GustafsonGreenland2010}). Thus, frequentist thinking can address Bayesian
overconfidence just as Bayesian thinking can address frequentist
overconfidence.\break Hence I would strengthen Box's plea for ecumenism (Box, \citeyear{Box1983}) into an imperative to fuse Bayesian and frequentist concepts and
methods in statistical\break inference---and in teaching as well. This theme
is far from new (e.g., besides Box, see Good, \citeyear{Good1983};
Diaconis and Freedman and discussants, \citeyear{DiaconisFreedman1986};
Samaniego and Reneau, \citeyear{SamaniegoReneau1994}), yet it has
barely touched everyday teaching and practice. In this case (unlike
many) that is not because of software limitations; in fact, for the bulk
of applications the same software can be used for both frequentist and
Bayesian calculations (Greenland, \citeyear{Greenland2007,Greenland2009a}).

%s1 ###
\section*{Hierarchical Modeling: Where Priors and Frequencies Meet}

Bayesian and frequentist ideas intertwine in hierarchical modeling
(Efron's Section~9), which encompasses both Bayesian and empirical-Bayes
approaches (Good, \citeyear{Good1983,Good1987}) as well as other shrinkage techniques.
Efron and Morris (\citeyear{EfronMorris1973,EfronMorris1975}) were among the earliest to demonstrate
convincingly that hierarchical models offered practical as well as
theoretical advantages for data analysis. Their writings (along with
those of Jack Good, George Box and Edward Leamer) inspired my
applications of hierarchical modeling and Bayesian methods in
epidemiology, where the hierarchy levels are naturally determined by
physical structures and observation processes.

As an example, in nearly all observational studies of nutrient effects,
individual risks are regressed directly on nutrient intakes calculated
from food intakes. This conventional model makes no further use of the
food intakes, and so assumes implicitly that foods have no effect on
risk beyond their calculated nutrient content. This is an unsupported
and very doubtful assumption. A more realistic model allows food effects
beyond measured-nutrient content. However, the resulting two-level
hierarchical model is not identified without a prior because nutrient
intakes are linear functions of food intakes (making nutrient and food
intakes completely collinear). Using any contextually defensible prior
reveals that the conventional analysis generates overconfident
inferences, both in the Bayesian sense of overstating information
(Greenland, \citeyear{Greenland2000}), and also in the frequentist sense of producing
interval undercoverage (Gustafson and Greenland, \citeyear{GustafsonGreenland2006}). That
overconfidence may explain the rather embarrassing track record of
nutritional epidemiology when compared against clinical trials (Lawlor et al., \citeyear{LawlorEtAl2004}).
Ecologic analyses provide other examples in which use of
the natural hierarchical structure with explicit priors is needed to
avoid overconfidence (Wakefield, \citeyear{Wakefield2009}).

In this work, I have come to appreciate that a simultaneously Bayesian
and frequentist viewpoint is essential for a credible analysis of
observational data. I must be at least informally Bayesian, knowing that
there is no contextual credibility without consideration and use of
prior information, especially in model specification. But I should also
be at least informally frequentist, knowing that priors should be
weighted lightly unless they derive from statistical observations such
as frequencies in partially exchangeable past experience (e.g., surveys)
or classical measurement processes (e.g., laboratory determinations).
Most of all, I should not rigidly adhere to ideologies or models,
especially when a clash between my prior and my likelihood function
shows that my understanding of the situation is more deficient than I
initially thought (Box, \citeyear{Box1980,Box1990}).

%s2 ###
\section*{Priors: Everybody Uses Them (but Most Call Them ``Models'')}

As Efron illustrates in Section 4, all analyses labeled as frequentist
are built on priors, although these priors are called ``models,'' which
avoids the controversies associated with overtly Bayesian analysis
(Leamer, \citeyear{Leamer1978};
Box, \citeyear{Box1980,Box1983}). Even the simplest randomized-trial
analysis is based on a model, namely the prior belief that treatment was
randomized fairly, and the reported subjects actually exist. As numerous
cases of fraud demonstrate, that belief may be mistaken more often than
those receiving medical treatment would like to think (e.g., see
Greenland, \citeyear{Greenland2009b}).

Labeling assumptions and models as prior beliefs might better alert us
to the act of faith involved in their use. As Box (\citeyear{Box1980}) said

\begin{quotation}
I believe that it is impossible logically to distinguish between
model assumptions and the prior distribution of the parameters. The
model is the prior in the wide sense that it is a probability statement
of all the assumptions currently to be tentatively entertained a priori.
On this view, traditional sampling theory was of course not free from
assumptions of prior knowledge. Instead it was as if only two states of
mind had been allowed: complete certainty or complete uncertainty.
\end{quotation}

I have grown increasingly uncomfortable with the convention of failing
to label models as priors. It encourages the use of arbitrary
constraints, and questions constraints only if the analysis data (the
direct evidence) can reveal departures---even though studies are not
designed with anywhere near sufficient power to reveal all important
model violations. The representation of modeling constraints in belief
networks (Madigan, Mosurski and Almond, \citeyear{MadiganMosurskiAlmond1997}) can aid in the display of these
constraints as imposed beliefs and thus expose implausible aspects of
the model, although of course it cannot address data limitations. Yet
single datasets are often too limited to tell us much about either the
effects under study or our models (Robins and Greenland, \citeyear{RobinsGreenland1986})---at
least if we do not impose a hoard of dubious independence constraints
that amount to point-mass priors with no supporting data.

Additivity in generalized linear models is an example: with $n$
covariates, additivity sets all orders of product terms
(``interactions'') among them to zero, and is equivalent to using a
point mass at zero for the joint prior on these terms. Entering the few
``significant'' two-way products hardly makes a dent in this set of
constraints if $n > 5;$ yet $n > 8$ is common and $n > 20$ not unusual.
Arbitrary additive constraints can be relatively harmless when
estimating a population-average effect, because the specification error
they entail may average out in much the way random residual error does
(Greenland and Maldonado, \citeyear{GreenlandMaldonado1994}). But the constraints can be deadly when
used for individual (clinical) risk prediction, as adverse drug
interactions demonstrate.

Hierarchical methods offer one way to relax additivity constraints in a
controlled fashion, by including all or many products but shrinking
their estimates toward zero or a second-level structure
(Wakefield, De Vocht and Hung, \citeyear{WakefieldDeVochtHung2010}). More generally, we can expand an unrealistic conventional
model by embedding it in a richer, more realistic hierarchical model,
then shrink estimates from the latter using prior distributions. Aspects
of these distributions may be chosen to improve frequency performance in
high-dimensional problems, but such methods do not preclude the use of
prior information to judge those and other aspects of the formal prior
distribution.

%s3 ###
\section*{The Need for Explicit Priors in Observational Studies}

My discomfort with conventional treatments of modeling has increased
knowing that observational data analysis can identify causal effects
\textit{only} by using indirect evidence, no matter how large the
dataset or how informed by past observational data. This is the usual
situation in epidemiology, where confounders, selection-probability
ratios, or valid exposure measurements are unavailable for analysis
(Greenland, \citeyear{Greenland2005a};
Gustafson, \citeyear{Gustafson2005};
Rothman,\break Greenland and Lash, \citeyear{RothmanGreenlandLash2008}, Chapter~19;
Lash, Fox and Fink, \citeyear{LashFoxFink2009}). The problem is a variant of the nonidentifiability of a
regression coefficient when some regressors are latent (Leamer, \citeyear{Leamer1974}).
In these cases a credible formal analysis must introduce proper priors
in place of overconfident identifying constraints.

Use of identified regression models as sources of effect estimates
corresponds to a multidimensional point prior that says there is no
uncontrolled confounding or selection bias, and that measurements
(including validation measurements) were accurate or at least reliable
for life histories. Taken jointly, these assumptions are absurd in
topics like nutritional and ``lifestyle'' epidemiology. But
relaxing\break
these silly and harmful assumptions leads to a realm where most
Bayesians as well as frequentists fear to tread: Specification of prior
distributions that cannot be effectively checked or updated with the
analysis data.

When the scientific validity of each analysis hinges on extensive and
untestable prior specifications, an analysis can be no more than a rough
guess about a vast unknown, and represents but one element in a
sensitivity analysis (Greenland, \citeyear{Greenland2005b}). This is true even of a formal
sensitivity analysis, which is limited to examining a few parameters
lest it become unintelligible. In this reality, the importance of
specific models and priors should be de-emphasized in favor of providing a
framework for sensitivity analysis across plausible models and priors.
Accuracy of computation becomes secondary to prior specification, which
is too often neglected under the rubric of ``objective Bayes'' (a.k.a.
``please don't bother me with the science'' Bayes).

There is simply no point in trying to do well at all conceivable
parameter values given the model when the model embedding the parameter
has already imposed doubtful point constraints. Hence I have sought
approaches in which informative priors are central. Good (\citeyear{Good1983}) provided
the key ingredients: Priors can be transformed into penalty functions,
which can then be transformed into ``prior data'' that generate the
penalties as log-likelihood contributions. This transformation allows
evaluation of prior-knowledge claims in a currency familiar to the
subject-matter expert, as well as use of familiar and rapid fitting
methods for basic models (Bedrick,\break Christensen and Johnson, \citeyear{BedrickChristensenJohnson1996,BedrickChristensenJohnson1997};
Greenland, \citeyear{Greenland2006,Greenland2007,Greenland2009a,Greenland2009c}).

Note that conversion of priors to prior data does \textit{not} require
conjugacy; it only requires that the penalties have representations as
transformed likelihoods from a series of observations or experiments.
The credibility of the prior may be questioned if such a representation
is absent, arcane, or absurd. Evaluation of priors in terms of
equivalent data is particularly illuminating in human-subject fields,
where data are expensive and hence sparse. Here, strong priors may be
seen as claiming access to a volume of data that does not exist, thus
casting doubt on prior assertions of some experts
(Higgins and Spiegelhalter, \citeyear{HigginsSpiegelhalter2002};
Greenland, \citeyear{Greenland2006}).

When priors (the indirect evidence) are recalibrated to match the
frequentist outputs of reasonably sized thought experiments, the
combined evidence will often be too limited to distinguish among the
effect sizes at issue (Greenland, \citeyear{Greenland2009c}). This is unwelcome news to some
colleagues, albeit no news to others. Regardless, the future of indirect
evidence should be recognition for what it is: Omnipresent and essential
for any inference beyond ``more research is needed'' (which may the
strongest conclusion we can hope to wrest from most studies, albeit not
always justifiable in economic terms).

Thus I would conclude by echoing Efron: Whether Bayesian, frequentist,
ecumenic, or syncretic, statisticians need to become better at creating
and evaluating contextually informed models---which include both
well-informed prior distributions and sensible qualitative structures.
It follows that statistical training should introduce
informative-Bayesian methods in tandem with classical (and often
destructive) frequentist methods, rather than as an afterthought or
specialty topic. Data priors provide one easy and natural way to do so,
displaying as they do the symmetry between indirect and direct evidence,
and exposing priors to a new angle of criticism.

\end{document}